# About the Number of Base Substitutions Between Humans and Common Chimpanzees

**Viswanath.C.Narayanan**

**Govt. Engineering College, Thrissur, India**

**Email: narayanan_viswanath@yahoo.com**

**Abstract**

Humans and chimpanzees are believed to have shared a common ancestor about 6 million years ago. Here using a new distance measure called the Jump distance, we calculate the number of base substitutions that might have occurred in the mitochondrial DNA during these 6 million years.

## 1.Introduction

Because of the rapid rate of changes, Mitochondrial DNA has got its own place in phylogenetic analysis. In the literature, different substitution rates for mitochondrial bases ranging roughly 0.025-0.26/Site/Myr (Parsons et. al. [1]), are taken for phylogenetic study. On the basis of different studies, it is believed that humans and chimpanzees were diverged from some common ancestor about 6 million years ago. Here we conduct a distance-based analysis to find the approximate number of mitochondrial base substitutions that might have occurred between these two species during a period of 6 million years. For this we assume an average base change rate of 0.15/Site/Myr.

Distance metrics plays an important role in phylogenetic reconstruction. We have various distance metrics used in phylogenetic analysis which are based on different DNA substitution models such as: Jukes-Cantor [2], Kimura [3,4], Felsenstein [5,6], Hasegawa, Kishino and Yano [7,8], Tamura and Nei [10], Posada [11] and Tavare [9]. In each of these models, we have a base substitution process which is a continuous-time Markov chain with states $\{A,C,G,T\}$, a 4×1 vector of equilibrium probabilities $\pi$ and a 4×4 rate matrix Q. Among all these models the GTR (generalized time reversible) model by Tavare [9] is the most general model in the sense that the rate matrix Q for these model generalizes the rate matrices for the other models. For a more detailed discussion see Huelsenbeck et al. [12]. There are distance metrics under more complex models, which can treat the case of unequal evolutionary rates across lineages, like the one discussed in Galtier and Gouy [13], the paralinear distance [14] and the LogDet distance [15,16].

Very recently Minin and Suchard [17], studied how to count the transitions in an evolutionary Markov model and based on this, O'Brien, Minin and Suchard [18], introduced a new method called robust counting that can be applied to a standard evolutionary model like the F84 model, for getting a better distance measure called the robust distance.

In Viswanath [19], we defined a new distance function by assuming that given two DNA sequences X and Y, during evolution of the sequence Y from sequence X (or vice versa), each base undergoes changes that determined by a continuous-time Markov chain with state space $\{A,C,G,T\}$ and infinitesimal generator matrix Q. Counting the number of transitions in the underlying Markov Chain, we defined a distance function (see [19]) which will be called as the Jump distance here. The jump distance was denoted as $d_Q(t)$ indicating it is the average number of changes that occurred to a base, which is subject to changes that are driven by a Markov chain with generator matrix Q, in the interval (0, t].

In this paper, we use the jump distance to calculate the number base changes that might have occurred between humans and chimpanzees.

## 2. Materials and Methods

### 2.1 Materials

We selected the mitochondrial DNA sequences of Homo sapiens (NC 012920) and Pan troglodytes (NC 001643) from the GenBank database. These sequences were them aligned using ClustalW [20]. The jump distance between humans and chimpanzees are then calculated by assuming the following forms for the Q matrix.

### 2.2 The different substitution models used

We calculated the jump distance $d_Q(t)$ assuming 4 different models. The first three definitions were based on the Jukes-Cantor, F84 and Kimura-2-parameter models respectively. The generator matrix Q in these three cases were named as $Q_J, Q_F$, and $Q_K$ respectively. These matrices are given by:

$$Q_J = \begin{bmatrix} * & \mu & \mu & \mu \\ \mu & * & \mu & \mu \\ \mu & \mu & * & \mu \\ \mu & \mu & \mu & * \end{bmatrix} \quad Q_F = \begin{bmatrix} * & \pi_A & \pi_C & \pi_G \\ \pi_T & * & \pi_C & \pi_G \\ \pi_T & \pi_A & * & \pi_G \\ \pi_T & \pi_A & \pi_C & * \end{bmatrix} \quad Q_K = \begin{bmatrix} * & \alpha & \beta & \alpha \\ \alpha & * & \alpha & \beta \\ \beta & \alpha & * & \alpha \\ \alpha & \beta & \alpha & * \end{bmatrix}$$

A fourth model with the following Q matrix is also studied:

$$Q_D = \begin{bmatrix} * & ac & ag & at \\ ca & * & cg & ct \\ ga & gc & * & gt \\ ta & tc & tg & * \end{bmatrix}$$

For easy identification of each of these definitions, let us rename $d_Q(t)$ in each of these cases as $d_{QJ}(t)$, $d_{QF}(t)$, $d_{QK}(t)$ and $d_{QD}(t)$ respectively.

For finding the total number of base substitutions that might have happened between human and chimpanzee mitochondrial genome, we make the following assumptions:

1. Humans and Chimpanzees diverged from a common ancestor around 6 million years ago.
2. Different base changes occur according to the particular rate matrix Q with parameters (base change rates per one million year) as given in table 1.

According to the definition, the distance function $d_Q(t)$ gives the average number of changes occurred to a base in the time interval (0, t], assuming that the changes are driven by a Markov chain with generator matrix Q. Since we want to measure the number of changes during 6 million years, we fix t = 6. Now agreeing that there are 16000 bases in the mitochondrial DNA, we calculate the total number of Mitochondrial base changes as equal to 16000 x $d_Q(6)$.

## 3  Numerical results

We fixed the average rate of base change per one million year as 0.15 in all the substitution models studied. The jump distance was found the highest in the case when the substitution model was taken as Kimura - 2 - parameter model. In the case of the other three models, the distance values were not too far from one another. The average jump distance was found to be 0.681 and the average number of base substitutions between Humans and Chimpanzees was obtained as 10899. Table 2 contains the number of base substitutions for different models of substitution process.

## 4. Tables and Figures

| Distance | Parameters | Average rate |
|---|---|---|
| $d_{QJ}(t)$ | $\mu = 0.15$ | 0.15 |
| $d_{QF}(t)$ | $\pi_A = 0.12, \pi_C = 0.13, \pi_G = 0.17, \pi_T = 0.18$ | 0.15 |
| $d_{QK}(t)$ | $\alpha = 0.2, \beta = 0.1$ | 0.15 |
| $d_{QD}(t)$ | $ac = 0.11,\ ag = 0.16,\ at = 0.18,\ ca = 0.14,$ $cg = 0.13,\ ct = 0.18,\ ga = 0.12,\ gc = 0.13,$ $gt = 0.2,\ ta = 0.15,\ tc = 0.19,\ tg = 0.11$ | 0.15 |

**Table 1:** Parameters for simulation experiments

| Distance function | | Number of base Substitutions $=16000 \times d_Q(t)$ |
|---|---|---|
| $d_{QJ}(t)$ | 0.659 | 10551 |
| $d_{QF}(t)$ | 0.661 | 10578 |
| $d_{QK}(t)$ | 0.74 | 11848 |
| $d_{QD}(t)$ | 0.664 | 10619 |

**Table 2:** The table gives the approximate average number of Mitochondrial base changes that might have occurred between humans and common chimpanzees.

## References

1. Parsons TJ et al, A high observed substitution rate in the human mitochondrial control region, Nat Genet 15, 363-368

2. Jukes, T.H, and C.R. Cantor. (1969). Evolution of protein molecules. Pp. 21-123 in H.N. Munro, ed. Mammalian Protein Metabolism. Academic Press, New York.

3. Kimura, M. (1980). A simple method for estimating evolutionary rates of base substitutions through comparative studies of nucleotide sequences. J. Mol. Evol. 16: 111-120.

4. Kimura, M. (1981). Estimation of evolutionary distances between homologous nucleotide sequences. Proc. Natl. Acad. Sci. USA. 78: 454-458

5. Felsenstein, J. (1981). Evolutionary trees from DNA sequences: a maximum likelihood approach. J. Mol. Evol. 17: 368-376

6. Felsenstein, J. (1984). Distance methods for inferring phylogenies: a justification. Evolution. 38: 16-24

7. Hasegawa, M., T. Yano and H. Kishino (1984). A new molecular clock of mitochondrial DNA and the evolution of Hominoids. Proc. Jpn. Acad. Ser. B 60: 95-98.

8. Hasegawa, M., H. Kishino and T. Yano (1985). Dating human-ape splitting by a molecular clock of mitochondrial DNA. J. Mol. Evol. 22: 160-174.

9. Tavare, S. (1986). Some probabilistic and statistical problem on the analysis of DNA sequences. Pp. 57-87 in Lectures in Mathematics in the Life Sciences, Vol. 17.

10. Tamura, K. and M. Nei (1993). Estimation of the number of nucleotide substitutions in the control region of mitochondrial DNA in humans and chimpanzees. Mol. Biol. Evol. 10: 512-526

11. Posada D. (2003). Using Modeltest and PAUP[*] to select a model of nucleotide substitution. Pp. 6.5.1-6.5.14 in A.D. Baxevanis, et al. eds. Current Protocols in Bioinformatics. John Wiley & Sons, Inc., New York.

12. John P. Huelsenbeck, Bret Larget and Michael E. Alfaro. (2004). Bayesian phylogenetic model selection using reversible jump Markov chain Monte Carlo. Mol. Biol. Evol. 21(6): 1123-1133.

13. Galtier. N, and Gouy. M. (1995). Inferring phylogenies from DNA sequences of unequal base compositions. Proc. Natl. Acad. Sci. USA 92: 11317-11321.

14. Lake, J.A. (1994). Reconstructing evolutionary trees from DNA and protein sequences: paralinear distances. Proc.Natl.Acad.Sci. USA 91:1455-1459.

15. Steel, M.A. (1994). Recovering a tree from the leaf colourations it generates under a Markov model. Appl.Math.Lett. 7:19-24

16. Lockhart, P.J., M.A. Steel, M.D. Hendy, and D. Penny. (1994). Recovering evolutionary trees under a more realistic model of sequence evolution. Mol.Biol.Evol. 11: 605-612.

17. V N Minin and M A Suchard (2008). Counting labeled transitions in continuous-time Markov models of evolution. J. Math. Biol 56: 391-412.

18. J D O'Brien, V N Minin and M A Suchard (2009). Learning to count: Roubust estimates for labeled distances between molecular sequences. Mol. Biol. Evol. 26(4): 801-814

19. Viswanath C.N (2009). A new distance between DNA sequences, arXiv:0902.1821v1[q-bio.PE].

20. Chenna, Ramu, Sugawara, Hideaki, Koike,Tadashi, Lopez, Rodrigo, Gibson, Toby J, Higgins, DesmondG, Thompson, Julie D. (2003) Multiple sequence alignment with the Clustal series of programs. *Nucleic Acids Res 31 (13):3497-500 PubMedID: 12824352*